\documentclass[11pt,a4paper]{article}
\pdfoutput=1
\usepackage{jheppub}
\usepackage[T1]{fontenc}
\usepackage[latin9]{inputenc}
\usepackage{graphicx}
\usepackage{tabularx}
\usepackage{esint}

\newcommand{\be}{\begin{equation}}
\newcommand{\ee}{\end{equation}}
\newcommand{\bea}{\begin{eqnarray}}
\newcommand{\eea}{\end{eqnarray}}

\newcommand{\eqn}[1]{(\ref{#1})}

\newcommand{\eps}{{\cal E}}
\newcommand{\fig}[1]{Fig.~\ref{#1}}

\newcommand{\mt}[1]{\textrm{\tiny #1}}
\def\nc {N_\mt{c}}

\newcommand{\sac}{\, , \qquad}

\newcommand{\uh}{u_\mt{H}}

\newcommand{\cf}{{\cal F}}
\newcommand{\cb}{{\cal B}}
\newcommand{\ch}{{\cal H}}

\newcommand{\thy}{t_\textrm{hyd}}

\newcommand{\eq}[1]{(\ref{#1})}

\title{Holographic heavy ion collisions with baryon charge}
\author[a,b]{Jorge Casalderrey-Solana,}
\author[a,c]{David Mateos,}
\author[d]{Wilke van der Schee} 
\author[a]{and Miquel Triana} 
\affiliation[a]{Departament de F\'\i sica Qu\`antica i Astrof\'\i sica \&  Institut de Ci\`encies del Cosmos (ICC), Universitat de Barcelona, Mart\'{\i}  i Franqu\`es 1, 08028 Barcelona, Spain}
\affiliation[b]{Rudolf Peierls Centre for Theoretical Physics, University of Oxford, 1 Keble Road, Oxford OX1 3NP, United Kingdom}
\affiliation[c]{ICREA, Passeig Llu\'\i s Companys 23, 08010 Barcelona, Spain} 
\affiliation[d]{Center for Theoretical Physics, MIT, Cambridge, MA 02139, USA}

\date{\today}

\abstract{
We numerically simulate collisions of charged shockwaves in Einstein-Maxwell theory in anti-de Sitter space as a toy model of heavy ion collisions with non-zero baryon charge. The stress tensor and the baryon current become well described by charged hydrodynamics at roughly the same time.
The effect of the charge density on generic observables is typically no larger than 15\%. 
We find significant stopping of the baryon charge and compare our results with those in heavy ion collision experiments.
}  
  

\emailAdd{jorge.casalderreysolana@physics.ox.ac.uk}
\emailAdd{dmateos@icrea.cat} 
\emailAdd{wilke@mit.edu} 
\emailAdd{mtriana@ffn.ub.edu} 
 
\begin{document}

\begin{flushright}
ICCUB-16-027
 \\
\end{flushright}

\maketitle
\setlength{\parskip}{8pt}


\section{Introduction}
\label{intro}

Holography has provided useful toy models with which to study the quark-gluon plasma (QGP) created in heavy ion collision experiments. In these models one considers a caricature of the incoming ions consisting of two gravitational shock waves moving at the speed of light towards each other  in anti-de Sitter space (AdS) \cite{Chesler:2010bi,Casalderrey-Solana:2013aba,Casalderrey-Solana:2013sxa,Chesler:2015wra}. The collision results in the formation and subsequent hydrodynamization of a strongly coupled QGP which can be studied from first principles (see e.g.~\cite{CasalderreySolana:2011us,Chesler:2015lsa} for recent  reviews). 

The dynamics of the gravitational field in AdS encodes only the dynamics of the stress tensor in dual gauge theory. In other words, the collisions considered in the references above result in the formation of a neutral plasma. However, the plasma created in heavy ion collisions certainly carries some non-zero baryon charge which increases in importance as the energy of the collision decreases. The goal of this paper is therefore to provide the first simulation of charged shock wave collisions in AdS in order to model the formation and  hydrodynamization of a QGP with a non-zero baryon charge. The conserved  baryon current is dual on the gravity side to a Maxwell field, so we consider collisions in Einstein-Maxwell theory.\footnote{Homogeneous relaxation in this theory  has been considered in \cite{Fuini:2015hba}.}

\section{The model}
We start with the Einstein-Maxwell action with a negative cosmological
constant
\begin{equation}
S=-\frac{1}{16\pi G}\int d^{5}x\sqrt{-g}\left(R+\frac{12}{L^{2}}-\frac{1}{4}e^{2}L^{2}F_{mn}F^{mn}\right)\,,
\label{action}
\end{equation}
where $G$ is Newton's constant, $R$ is the Ricci scalar, $L$
is the asymptotic AdS radius, $e$ is a parameter controlling the backreaction of the Maxwell field on the metric, and \mbox{$F_{mn}\equiv\partial_{[m}A_{n]}$} is the electromagnetic field strength with $A_{m}$ the  vector potential. The metric and the gauge field are respectively dual on the gauge theory side  to the  stress tensor $T_{\mu \nu}$ and to a conserved $U(1)$ current $J^\mu$ whose time component one is free to think of as the baryon number density. In the case in which the action \eq{action} is viewed as a consistent truncation of the dimensional reduction of type IIB supergravity on $S^5$ the dual gauge theory is $\mathcal{N}=4$ Super Yang-Mills and the $U(1)$ current arises from the R-symmetry of this theory.\footnote{The full five-dimensional action for this truncation would include a Chern-Simons term (see e.g.~\cite{Gauntlett:2003fk}), but this will play no role in our analysis and we have therefore omitted it.} When we need to be concrete (for example to fix normalization factors) we will adopt this viewpoint. However, we emphasize that for most purposes the specific origin of the Maxwell field is unimportant and one could think of \eq{action} simply as a bottom-up model that incorporates the minimal set of ingredients to describe the dynamics of the stress tensor and a conserved $U(1)$ current in the dual gauge theory.

The equations of motion following from \eq{action} are
\begin{eqnarray}
R_{mn}+\frac{4}{L^{2}}g_{mn} & = & e^{2}L^{2}T_{mn}\,,
\label{einstein} \\[1mm]
\partial_{m}\left(\sqrt{-g}F^{mn}\right) & = & 0,
\end{eqnarray}
where 
\be
T_{mn}=\frac{1}{2}F_{mp}F_{n}^{\,\,\,p}-\frac{1}{12}g_{mn}F^{2} 
\ee
is the stress tensor sourced by the electromagnetic field.  These equations admit the following solution describing a charged shock wave moving at the speed of light:
\begin{eqnarray}
ds^{2} & = & \frac{L^{2}}{u^{2}}\Big(-dx_{+}dx_{-}+dx_{\perp}^{2}+\Big[u^{4}h(x_{+})-\frac{1}{3}e^{2}u^{6}a(x_{+})^{2}\Big]dx_{+}^{2}+du^{2}\Big)\,,\label{eq:metric}\\
A & = & \frac{u^{2}}{L^{2}}\,a(x_{+})\,dx_{+}\,, \label{gaugefield}
\end{eqnarray}
where $x_{\pm}=t\pm z$ and  $h(x_{+})$ and $a(x_{+})$ are arbitrary
functions of $x_+$. In the case in which the dual theory is ${\cal N}=4$ SYM these functions are related to the expectation values of the corresponding dual operators through  \cite{deHaro:2000vlm}
\begin{eqnarray}
T_{++} & = & \frac{N_{c}^{2}}{2\pi^{2}}h(x_{+}) \,,\\
J_{+} & = & \frac{N_{c}^{2}e}{\pi^{2}}a(x_{+}) \,.
\end{eqnarray} 
The fact that $J$ scales as $\nc^2$ reflects its R-symmetry origin, namely that microscopically it is built out of adjoint degrees of freedom.  
An analogous solution describing a wave moving in the opposite direction is obtained by replacing $x_+ \to x_-$ in the expressions above.

We will adopt the following choices for the functions $h$ and $a$ describing our incoming projectiles:
\be
h(x_{\pm})=\frac{m^{3}}{\sqrt{2\pi w^{2}}} \, \exp \left(
-\frac{x_{\pm}^2}{2w^{2}}  \right) 
\sac a(x_{\pm}) = h(x_{\pm}) / 2m   \,.\\[2mm]
\label{initial}
\end{equation}
The energy and charge densities per unit area of the shock are   $2\pi^2 m^3/\nc^2$ and $2\pi^2 m^2/\nc^2$, respectively. Note that we choose the centers and the widths of the Gaussian profiles for $h$ and $a$ to be the same, as corresponds to the fact that we want to model the collision of projectiles that carry both energy and charge. 

We see from Eq.~\eq{einstein} that the parameter $e$ controls the magnitude of the backreaction of the Maxwell field on the dynamics of the spacetime metric. It is clear from the action \eq{action} that this parameter could be absorbed in the normalization of the Maxwell field, at the expense of including an explicit factor of $e$ multiplying the gauge field amplitude $a(x_{+})$ in Eq.~\eq{gaugefield}. In other words, one can either think of the backreaction as being controlled by $e$ for a fixed incoming amplitude $a(x_{+})$, or as it being controlled by the incoming amplitude for a fixed $e$. We find it convenient to adopt the first viewpoint. 

The limit $e\to 0$ on the gravity side corresponds to an approximation in which the Maxwell field is treated as a probe field that propagates in a fixed background without affecting it. In the gauge theory this means that the charge density is treated in a quenched approximation in which it has a negligible effect on the dynamics of the gluons. This limit is physically interesting since it describes the situation in which the energy density dominates over the charge density, and also as a benchmark against which the backreacted results can be compared. In Sec.~\ref{probe} we will consider the collision in the probe approximation, and  in Sec.~\ref{back} we will consider the backreacted case.    

\section{Thermodynamics and hydrodynamics}
\label{thermosec}
We begin by  reviewing the thermodynamics of the plasma in the theory described by the action \eq{action}. This will be useful when we later consider the hydrodynamics of the plasma. 

\begin{figure}[t]
\begin{center}
\includegraphics[width=8cm]{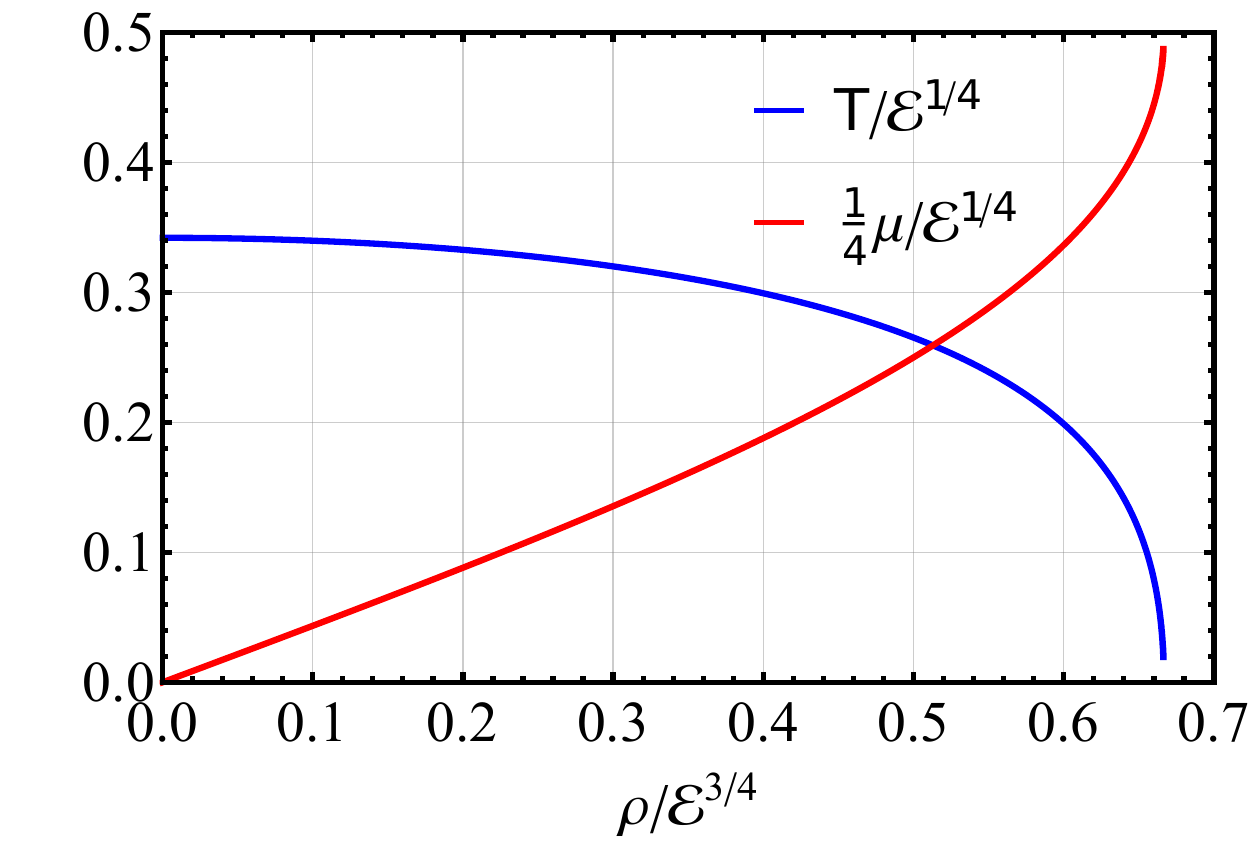}
\end{center}
\caption{Temperature and chemical potential as a function of the charge density, all normalized by the energy density.}
\label{eos}
\end{figure}

We will work with a rescaled stress tensor and current given by
\be
\mathcal{T}_{\mu\nu} = \frac{2\pi^2}{\nc^2} T_{\mu\nu} \sac
\mathcal{J}^\mu = \frac{2\pi^2}{\nc^2} J^\mu  \,,
\ee
and set
\be
\mathcal{E} = -  \mathcal{T}^0_0 \sac 
\rho= \mathcal{J}^0 \,.\label{eq:charge}
\ee
The relation between these quantities and the temperature $T$ and the chemical potential $\mu$ is given by the following expressions 
(see e.g.~\cite{Erdmenger:2008rm}):
\bea
\mathcal{E} &=& \frac{3}{4} \, x^4 \left(1+\sqrt{1+\frac{1}{6} \, y^2} \,\right)^3 
\left(3 \sqrt{1+\frac{1}{6} \, y^2}-1 \right) \,, \\[2mm]
\rho &=& \frac{1}{2} \, x^3 \, y  \left(1+\sqrt{1+\frac{1}{6} \, y^2} \,\right)^2 \,,
\eea
where
\be
x = \frac{\pi T}{2} \sac y = \mu / x\,.
\ee
Using these expressions one can show that, for a small charge density, the chemical potential is given by
\be
\mu \simeq \frac{\sqrt{3} \rho}{\sqrt{\mathcal{E}}} 
\left[ 1 + \frac{\sqrt{3}}{4} \left( \frac{\rho}{\mathcal{E}^{3/4}} \right)^2  + \cdots \right]\,.
\label{lin}
\ee
We see that, for fixed energy density, $\mu$ and $\rho$ are linear in one another with a 1\% (10\%) accuracy if $\rho/\mathcal{E}^{3/4}$ is no larger than 0.15 (0.48). In the opposite limit, when the chemical potential is high compared to the temperature, $y\gg 1$, we have that 
\be 
\eps \simeq \frac{\pi^4}{2^8} \, \mu^4 \sac
\rho\simeq \frac{\pi^3}{3 \cdot 2^5} \, \mu^3 \,,
\ee
and the ratio of charge to energy density approaches the value 
\be
\frac{\rho_\mt{max}}{\mathcal{E}^{3/4}} = \frac{2}{3} \,.
\label{ratio2}
\ee
The  subindex `max' indicates that this is the maximum value of $\rho$ for a fixed energy density. At this value the black brane becomes extremal and the temperature approaches zero compared to any other scale. Above this value a naked singularity appears. These features are illustrated in \fig{eos}.

Charged hydrodynamics for $\mathcal{N}=4$ SYM has been studied extensively
\cite{Erdmenger:2008rm,Banerjee:2008th}. In the case of a 1+1 dimensional flow, as is of interest here, the constitutive relations take the form\footnote{Note the the velocity curl term denoted $\ell_\mu$ in \cite{Erdmenger:2008rm} vanishes identically in 1+1 dimensions.}  
\begin{eqnarray}
\mathcal{T}_{\mu\nu} & = & \mathcal{E}\,u_{\mu}u_{\nu}+ 
P (\mathcal{E}) \Delta_{\mu\nu}- 
\eta \, \sigma_{\mu \nu}\,,
\label{eq:hydro-constituive} \\[2mm]
\mathcal{J}_{\mu} & = & \rho \,u_{\mu}-
\kappa \, \Delta_{\mu}^{\,\nu}\partial_{\nu} \left( \frac{\mu}{T} \right)\,,
\label{eq:Jhydro2nd}
\end{eqnarray}
where 
\begin{eqnarray}
\Delta_{\mu\nu} &=&  g_{\mu\nu}+u_{\mu}u_{\nu} \,, \\
\sigma_{\mu \nu}&=& \Delta^{\mu\alpha}\Delta^{\mu\beta} \left(   \nabla_{ \alpha} u_{\beta}   +   \nabla_{ \beta} u_{\alpha}\right) -\frac{2}{3} \Delta^{\mu \nu}\Delta^{\alpha \beta}  \nabla_{ \alpha} u_{\beta}   \,,
\end{eqnarray}
and the transport coefficients (rescaled by $2\pi^2/\nc^2$) are
\bea
\eta &=& \frac{1}{4} \, x^3 \left(1+\sqrt{1+\frac{1}{6} \, y^2} \,\right)^3 \,, \\[2mm]
\kappa &=& \frac{2}{\pi} \, x^2  \left(1+\sqrt{1+\frac{1}{6} \, y^2} \,\right) 
\left(3 \sqrt{1+\frac{1}{6} \, y^2}-1 \right)^{-1} \,.
\eea

\section{The probe approximation}
\label{probe}
With the initial conditions \eq{initial} for the collision 
we can solve the Einstein-Maxwell equations, as outlined in \cite{Chesler:2013lia,vanderSchee:2014qwa}.\footnote{The \emph{Mathematica} code to perform an evolution as described above is
available upon request at \href{mailto:wilke@mit.edu}{wilke@mit.edu};
alternatively, simpler versions can be found at \href{https://sites.google.com/site/wilkevanderschee}{sites.google.com/site/wilkevanderschee}}
In this section we work in the strict probe approximation for the Maxwell field. We are therefore computing the evolution of a Maxwell field on top of the dynamical shock wave collisions studied earlier in \cite{Chesler:2010bi,Casalderrey-Solana:2013aba,Casalderrey-Solana:2013sxa,Chesler:2013lia,vanderSchee:2014qwa,Chesler:2015fpa}. Computationally it is however convenient to evolve both the metric and the Maxwell field at the same time, thereby recomputing the gravitational shock wave background.

\begin{figure*}[t]
\begin{centering}
\includegraphics[width=13cm]{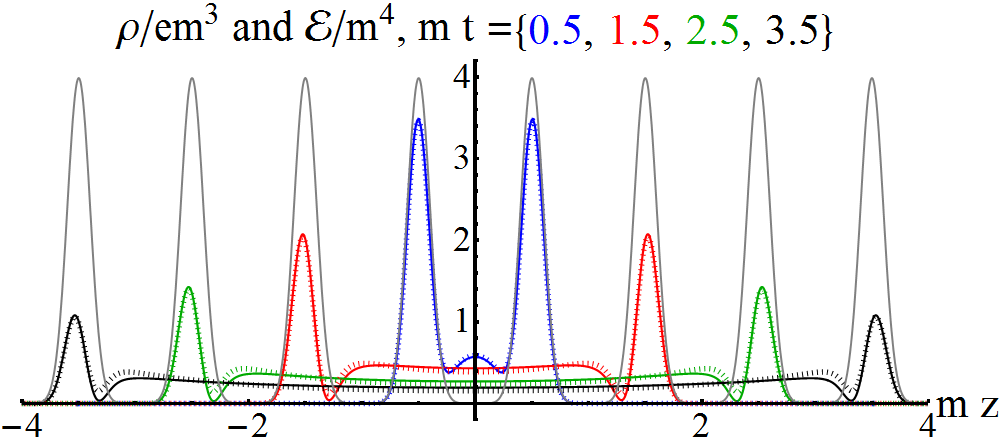}\\
\includegraphics[width=13cm]{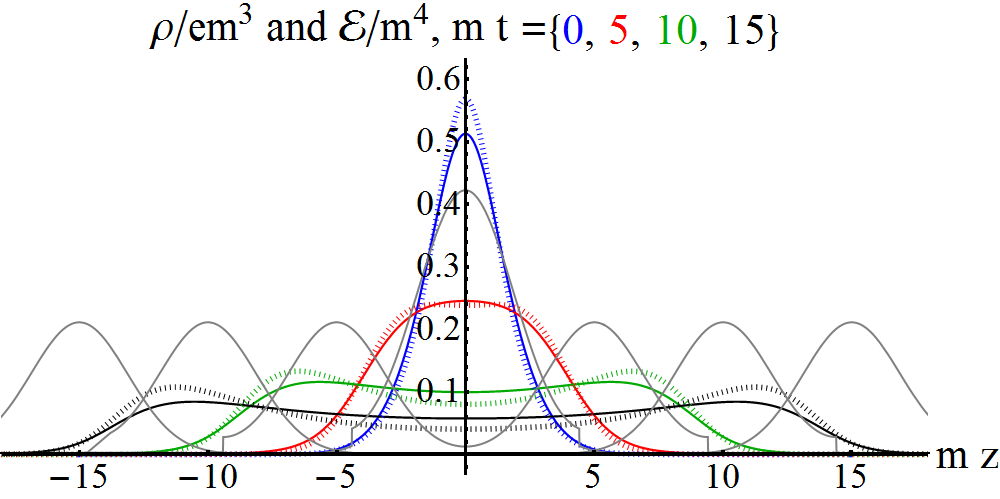}
\par\end{centering}
\caption{Snapshots of the charge density (solid curves) and energy density (dotted curves) for thin  (top) and thick (bottom) collisions. The grey curves indicate the charge density of the unperturbed original shocks. \label{fig:snapshots}}
\end{figure*}

We performed simulations for $m w = 0.1$ ($1/4\,$-shocks in the language of \cite{Casalderrey-Solana:2013aba}) and \mbox{$m w = 1.9$} ($2\,$-shocks in the language of \cite{Casalderrey-Solana:2013aba}), which we will refer to as thin  and thick shock collisions, respectively. The resulting charge density $\rho$ is plotted in \fig{fig:snapshots}  for both cases, 
whereby we included the energy density $\mathcal{E}$
for comparison (also found in \cite{Casalderrey-Solana:2013aba}). Clearly in the thin regime the shocks gradually lose their charge into a charged plasma between the shocks, much like the energy density. For the thick shocks the energy density already hydrodynamizes during the collision regime \cite{Casalderrey-Solana:2013aba} and, as we will see below, the same is true for the charge density. The shape of the charge and energy densities is different, a feature which we will later analyze in the local rest frame as well.

One interesting feature is the decay of the original shocks in the thin regime. Indeed, we see that on the attenuating maxima the charge density exactly follows the energy density, despite the fact that the charge and the energy are distributed differently in between the shocks. This agreement on the attenuating maxima is remarkable in view of the fact that the energy and the charge  are governed by in principle completely different dynamics, i.e.~the Einstein equations and the Maxwell equations on a fixed background, respectively. This suggests that on the light cone  a simplified picture may be possible (perhaps along the lines of \cite{Albacete:2009ji}).

Having the complete stress-tensor we are  able to extract the energy
and charge density in the local rest frame, defined through 
\be
T_{\mu}^{\nu}u_{\nu}=-\eps_\textrm{loc} \, u_{\mu} \,, 
\ee
from which we find
\bea
\eps_\textrm{loc}&=&\frac{1}{2} \left(\sqrt{(\mathcal{T}^{zz}+\mathcal{T}^{t t })^2-4 (\mathcal{T}^{tz})^2}-\mathcal{T}^{zz}+\mathcal{T}^{t t }\right) \,,
\nonumber \\[2mm]
v_\textrm{loc} &\equiv& \frac{u_z}{u_t} =\frac{\mathcal{T}^{zz}+\mathcal{T}^{t t }-\sqrt{(\mathcal{T}^{zz }+\mathcal{T}^{t t })^2-4 (\mathcal{T}^{t z })^2}}{2 \mathcal{T}^{t z }} \,,
\label{eloc}
\eea
where $u_{\mu}$ is the (timelike) fluid velocity. For the charge
density we have 
\be
\rho_\textrm{loc}=J_{\mu}u^{\mu} \,.
\ee
This allows us to study the approach to charged hydrodynamics. We read off from our simulations the fluid velocity and the energy  and charge densities.  We then use the constitutive relations \eq{eq:hydro-constituive}-\eq{eq:Jhydro2nd} to obtain the hydrodynamic prediction for the transverse and longitudinal pressures, $P_L^\textrm{hyd}$ and 
$P_T^\textrm{hyd}$, and for the time component of the current, $J_t^\textrm{hyd}$. We define the hydrodynamization time for the stress tensor, $\thy$, as the time beyond which 
\be
\frac{3 \left| P_L - P_L^\textrm{hyd} \right|}{\eps} < 0.1 \,.
\label{thyd}
\ee
Similarly, we define the hydrodynamization for the current, $\thy^J$, as the time beyond which
\be
\frac{\left| J_t - J_t^\textrm{hyd} \right|}{\rho} < 0.1 \,.
\label{thydj}
\ee
A crucial difference between the hydrodynamization of the stress tensor  versus the charge current is that the pressures at $z=0$ can deviate from their hydrodynamic  values due to the gradients of the velocity field. In contrast, parity symmetry implies that $u$ has no spatial component at $z=0$ and that  $J_z$ vanishes identically.  As a consequence, the (ideal) hydrodynamic prediction for the current is always exact at mid-rapidity. In order to asses the validity of the hydrodynamic description for $J$ we will therefore look at non-zero rapidity. 

\begin{figure*}[t]
\begin{centering}
\includegraphics[width=6.25cm]{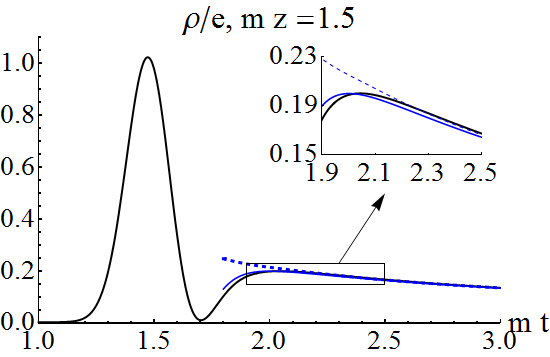}\,\,\includegraphics[width=7.75cm]{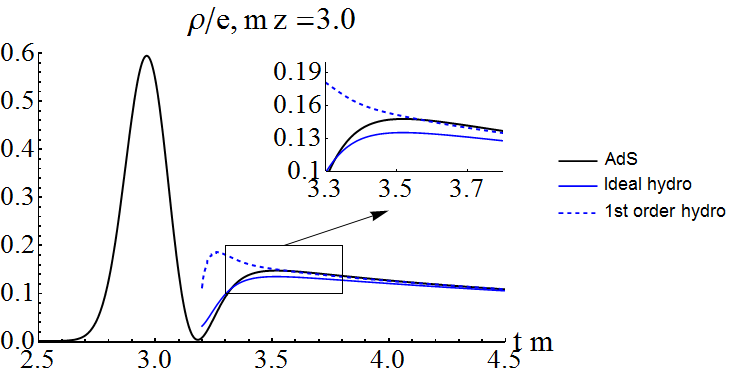}\\
\includegraphics[width=6.25cm]{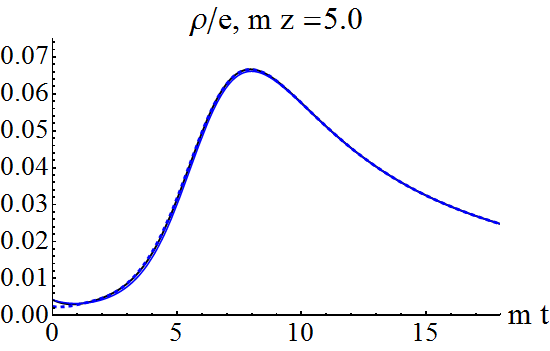}\,\,\includegraphics[width=7.75cm]{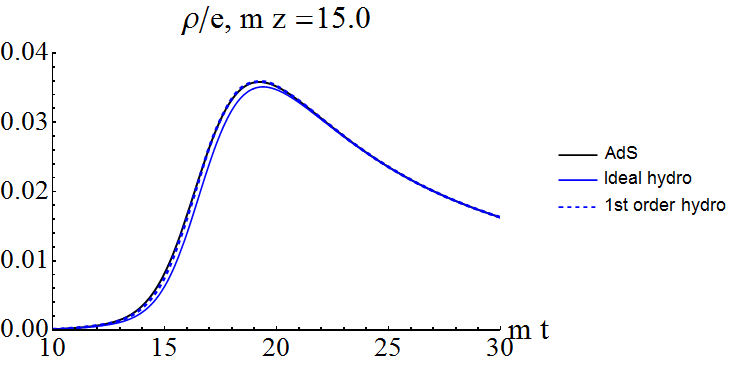}
\par\end{centering}
\caption{Time component of the current $J_t$ (black, solid curve) and its approximations based on ideal (blue, solid curve) and viscous (blue, dashed curve) hydrodynamics, for thin (top row) and thick (bottom row) shocks. The hydro curves start at a time after which a local rest frame can be defined (as in \cite{Arnold:2014jva}).
The insets show that even though ideal hydrodynamics gives a better overall fit in the range plotted, viscous hydrodynamics gives a better description of the final approach to hydrodynamics. Thick shocks (bottom) are always well described by charged hydrodynamics, whereby viscous 
hydrodynamics gives a significant improvement.
\label{fig:tohydro}}
\end{figure*}

The hydrodynamization times for the stress tensor in the case of a neutral fluid were determined in Refs.~\cite{Chesler:2010bi,Casalderrey-Solana:2013aba,Chesler:2015fpa}. The result 
%
is\footnote{Note that we have used a slightly different criterion  for hydrodynamisation than in  \cite{Casalderrey-Solana:2013aba}.} 
\bea
&& m \thy = \{ 2.0, 2.4, 3.6 \} \qquad \mbox{at} \qquad 
m z = \{ 0, 1.5, 3.0  \} \qquad \mbox{for thin shocks} \,, \\
&& m \thy = \{ 2.1, 4.7, 16 \} \qquad \,\, \mbox{at} \qquad 
m z = \{  0, 5, 15 \} \qquad \,\,\,\,\,\, \mbox{for thick shocks}\,.
\eea
Since in this section we are working in the probe approximation, these times agree with our  hydrodynamization times. With our new simulations we can now also study the hydrodynamization of the charge current, whose evolution and hydrodynamic approximations are shown in  \fig{fig:tohydro}.
The hydrodynamization times are 
\bea
&& m \thy^J = \{ 2.0, 3.5 \} \qquad \mbox{at} \qquad 
m z = \{ 1.5, 3.0  \} \qquad \mbox{for thin shocks} \,, \\
&& m \thy^J = \{ 0, 0 \} \qquad \,\, \,\,\,\,\,\,\,\, \mbox{at} \qquad 
m z = \{  5, 15 \} \qquad \,\,\,\,\,\, \mbox{for thick shocks}\,.
\eea
In the case of thick shocks we have listed the value $\thy^J=0$ to indicate that the current is always well predicted by hydrodynamics. In contrast, for thin shocks we see that the hydrodynamization times for the current away from mid-rapidity are very similar to those for the stress tensor.   In both cases, after $\thy^J$ there is a small but significant difference between ideal and viscous hydrodynamics, which shows that the fluid velocity of the charge differs from the velocity of the energy-momentum flow.

\begin{figure*}[t]
\begin{centering}
\includegraphics[width=7cm]{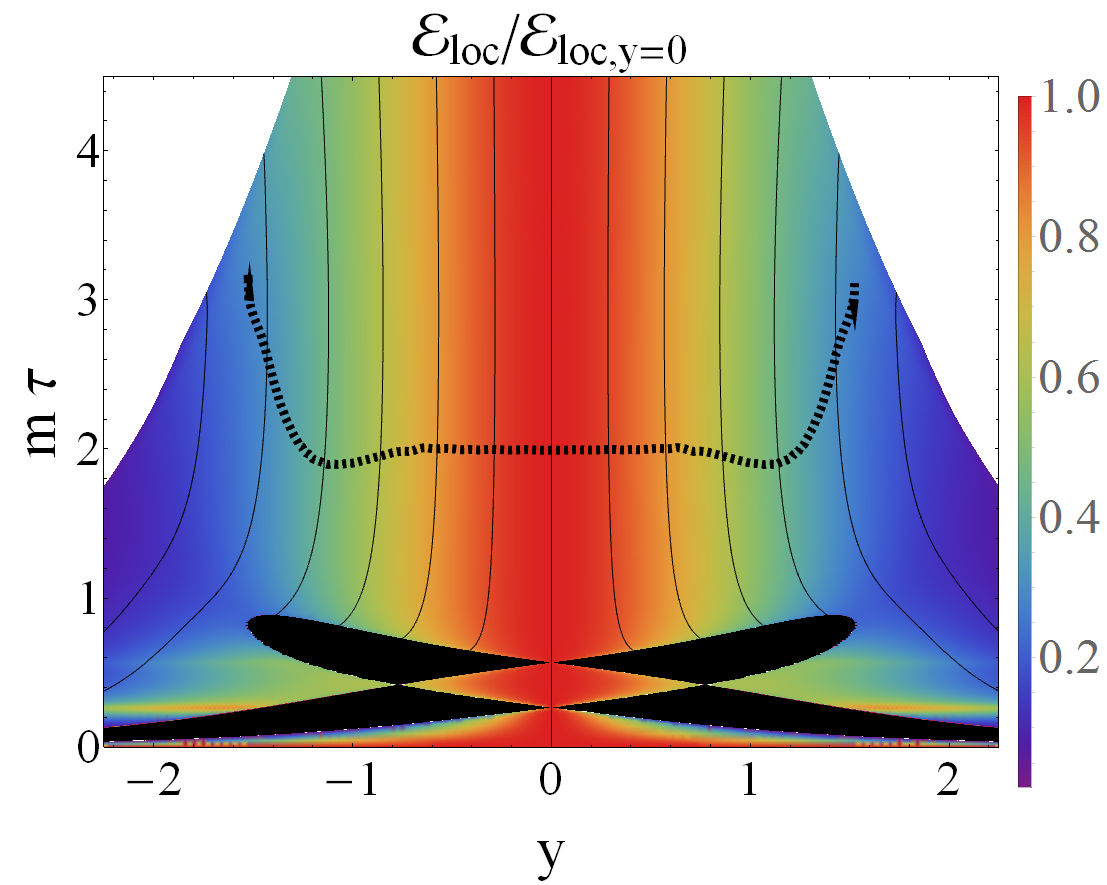}\includegraphics[width=7cm]{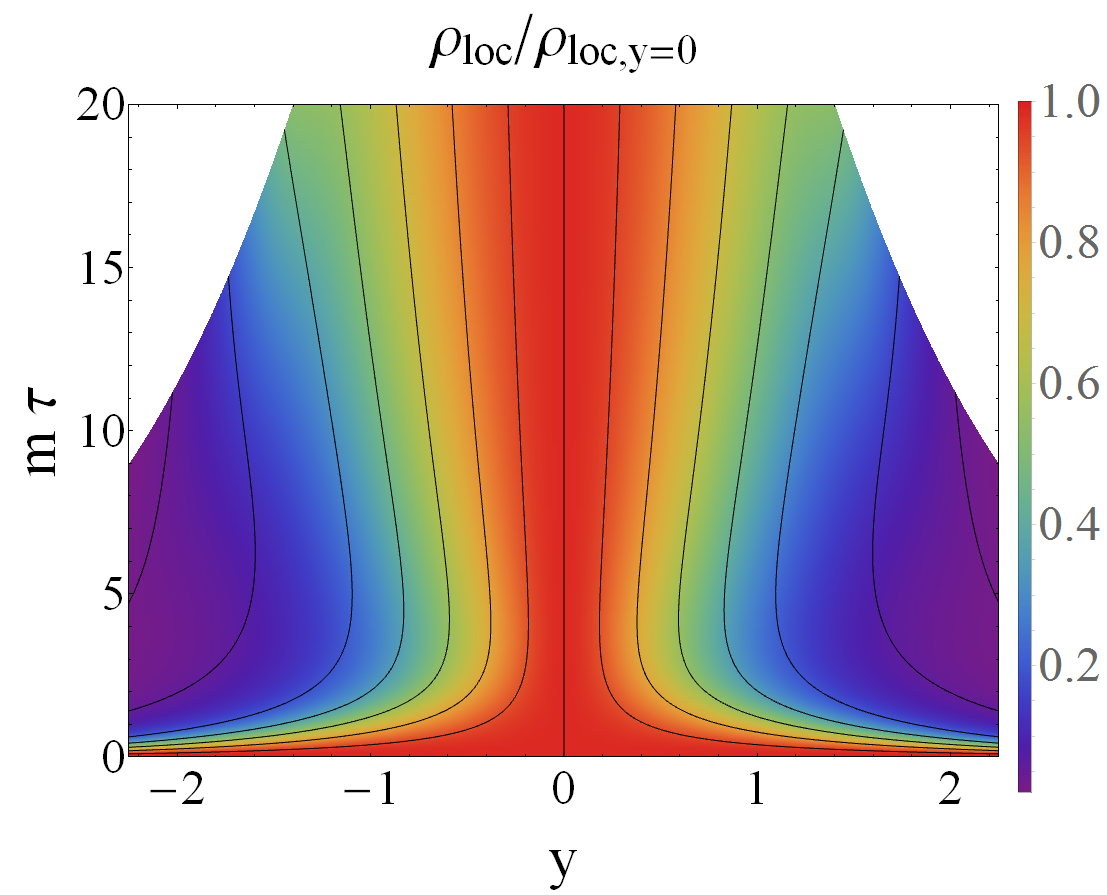}
\par\end{centering}
\caption{Local charge density
as a function of proper time $\tau$ and rapidity $y$. The left plot
shows the rapidity distribution for thin shocks, where regions without a rest frame are drawn black 
\cite{Arnold:2014jva},
and the hydrodynamic region is indicated by the dashed line (according to (\ref{thyd})). The upper white regions are outside our
numerical grid. The right plot is for thick shocks,
which is always in the hydrodynamic regime for $\tau>2$. The black
lines are stream lines of the fluid velocity, which are similar to the charge velocity.\label{fig:charge-rapidity}}
\end{figure*}

\begin{figure*}[t]
\begin{centering}
\includegraphics[width=7cm]{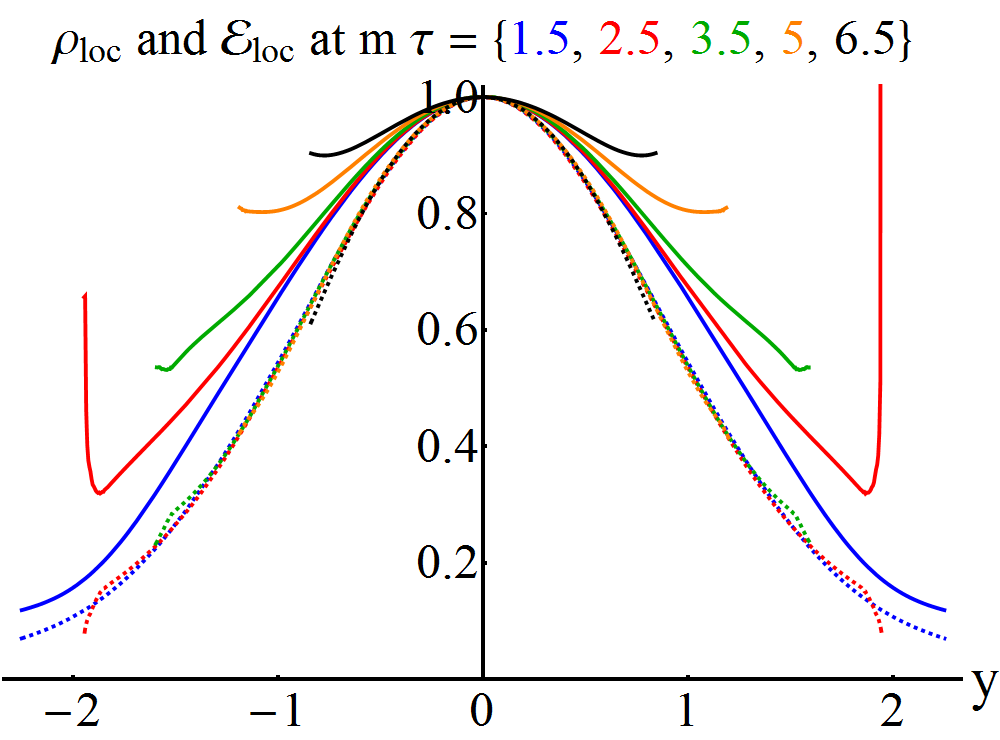}\includegraphics[width=7cm]{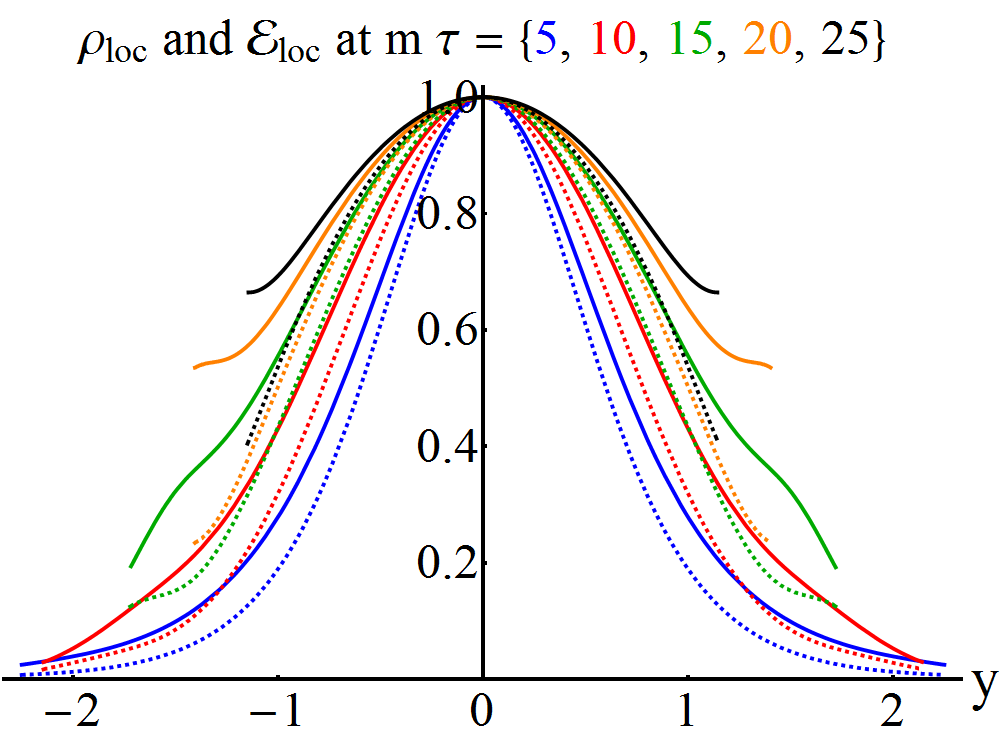}
\par\end{centering}
\caption{Rapidity profile of the local charge (solid curves) and the local energy  (dashed curves) densities at different proper
times $\tau$ as a function of rapidity $y$ and normalized at mid-rapidity (see also \cite{Casalderrey-Solana:2013aba,vanderSchee:2014qwa,Chesler:2015fpa}
for similar plots of the energy density). For thin shocks the energy density does not evolve much on the times shown, whereas the profile
for the charge density widens significantly. For thick shocks the charge and energy profiles widen in a similar fashion, though the profile of the charge is always wider.
\label{fig:energy-rapidity}}
\end{figure*}

\fig{fig:charge-rapidity} shows  the spacetime rapidity profile of the local charge density.  
This is compared with the local energy density rapidity profile for several different proper times in \fig{fig:energy-rapidity}. 

Especially striking in \fig{fig:energy-rapidity} is the development of maxima at non-zero rapidity for the charge profile; we will come back to this in the Discussion section. This happens fast in the case of thin shocks, whereas it takes longer for thick  shocks. Furthermore,
we see that for thin shocks the evolution of the charged rapidity profile is expanding much faster than the equivalent profile for the local energy density. For thick shocks both profiles  expand at a similar rate.

\begin{figure*}[t]
\begin{centering}
\includegraphics[width=7cm]{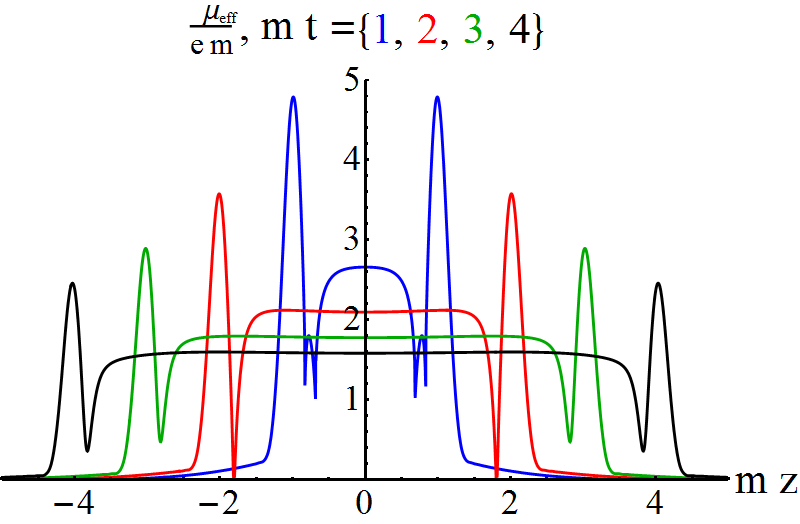} 
\,\,
\includegraphics[width=7cm]{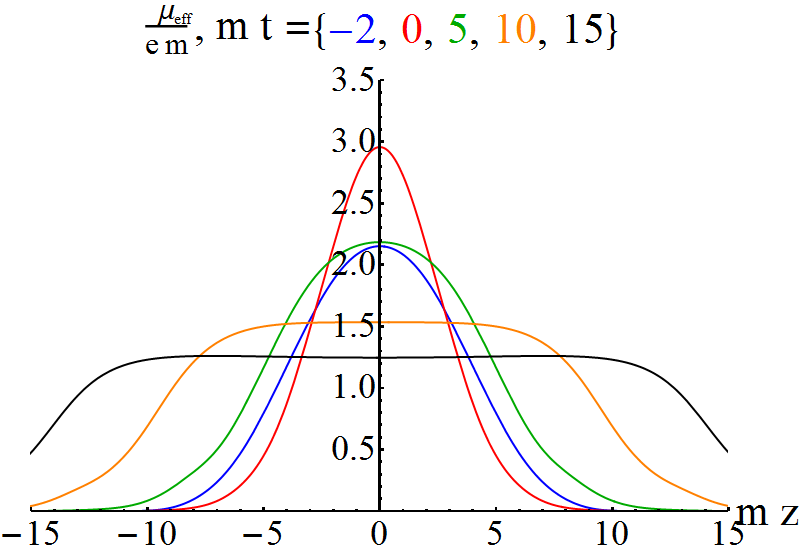}
\par\end{centering}
\caption{\small Chemical potential for thin (left) and thick  (right) shock collisions. 
\label{fig:chemical-potential}}
\end{figure*}

To characterize the charge deposition after the collision, we compare the charge density in the local rest frame, $\rho_\textrm{loc}$, to the local energy density, $\mathcal{E}_\textrm{loc}$. 
More precisely, at all points in spacetime at which a local rest frame exists, we define the ratio 
\be
\mu_\textrm{eff} \equiv \sqrt{3} \rho_\textrm{loc}/\mathcal{E}^{1/2}_\textrm{loc}\, .
\ee
The significance of this ratio, which we plot in \fig{fig:chemical-potential}, is that,  
in equilibrium, $\mu_\textrm{eff}$ coincides with the small-charge limit of chemical potential of the plasma, Eqn.~\eq{lin}. We will come back to \fig{fig:chemical-potential} in the Discussion section.
%
%

\begin{figure*}[t]
\begin{centering}
\includegraphics[width=9cm]{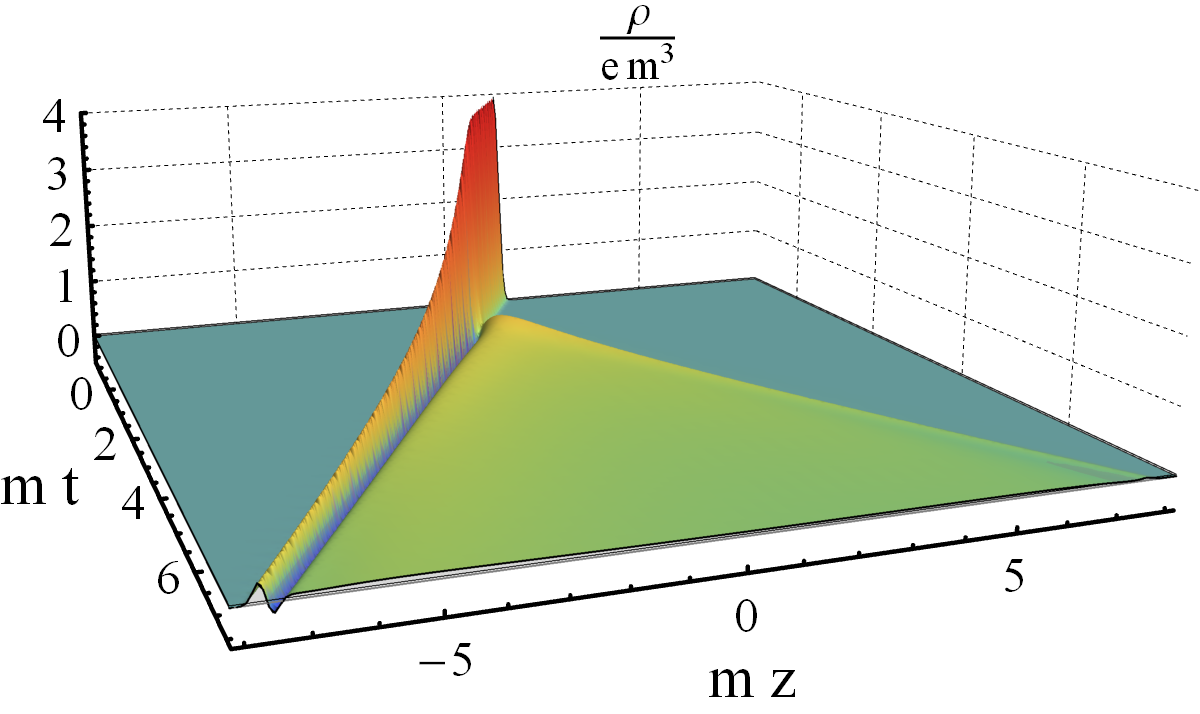}
\,\,
\includegraphics[width=5.5cm]{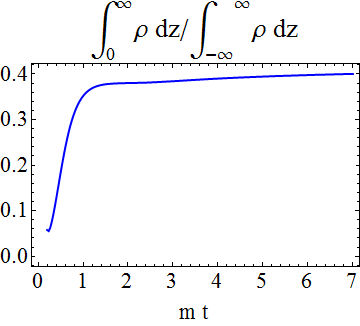}
 \\
\includegraphics[width=12cm]{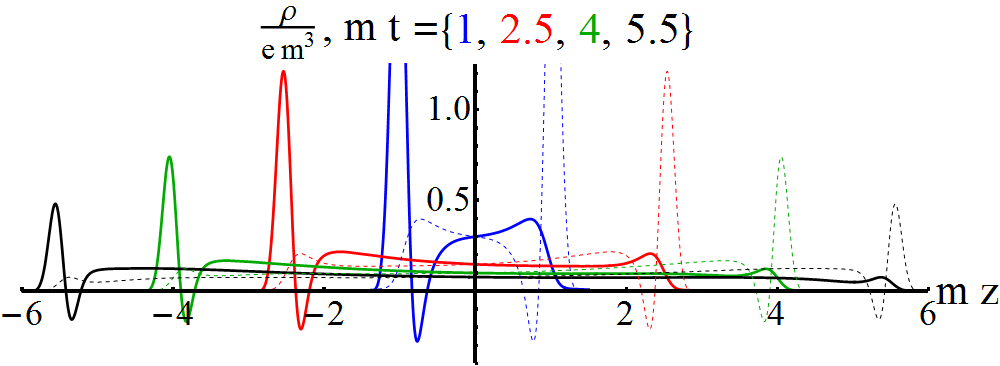} 
\par\end{centering}
\caption{\small We present the charge density $\rho$ for a $m w=0.1$ collision where only the left-moving shock is charged, with a 3D plot (top-left), snapshots with dashed lines showing a reflection (bottom) and with the fraction of the charge ending up at $z>0$ as a function of time (top-right). Interestingly, even though the initial charge moves towards negative $z$ at the speed of light the collision causes about 41\% ends up at positive $z$, indicating very strong interactions.
\label{fig:asymmetric-shocks}}
\end{figure*}

When colliding shocks with symmetric energy and charge distributions it is impossible to determine which part of the energy and charge in the final plasma comes from the left- or right-moving shock. In our current set-up, however, it is possible to charge only  the left-moving shock, leaving the right-moving shock neutral. This results in a charge distribution as shown in Fig.~\ref{fig:asymmetric-shocks}.\footnote{Note that without backreaction the charge of the colliding shocks does not interact with itself; the charge profile of the symmetric collision is hence equal to the sum of the dashed and solid lines in Fig. \ref{fig:asymmetric-shocks}.} Clearly most of the charge ends up at $z<0$, but at later times a surprisingly large fraction of about 41\% ends up at $z>0$, indicating that the strong interactions of the collision can indeed let the charge density bounce back, reflecting the direction of 41\% of the charge. 

Quite interestingly, the right-moving charge contains a bump moving close to the speed of light, though not as fast as the left-moving shock moving at the speed of light. This bump, however, does not come with a minimum with negative charge density, indicating that the negative charge density has to be associated to the presence of the original shock on the light cone. We verified that the profile away from the original shock is independent of the width of the charge, in agreement with the findings in \cite{Casalderrey-Solana:2013sxa}.
Experimentally, these simulations have potential consequences when it is possible to vary the baryon charge while leaving the energy constant. This can for instance be done by comparing proton with antiproton collisions, or deuteron collisions with proton collisions of double the energy (preferably selected such that the deuteron is aligned along the beam direction).

\section{Charge backreaction}
\label{back}
We now move away from the case $e=0$ and  include the full  backreaction of the charge density. We find that, for a given width of the incoming shocks, there is a maximum value of $e$ that our code is able to evolve. One clear physical reason why there is such a maximum value is that the charged plasma formed after the collision has to have a smaller charge density than the maximum density set by Eq.~(\ref{ratio2}), and indeed our simulations reach as high as 80\% of the maximum value according to the instantaneous energy density. On the other hand, this maximum value is also sensitive to the performance of the numerical code and furthermore depends sensitively on the initial separation of the shocks. The latter may be an indication that single charged shock waves cannot be evolved in a stable manner by themselves. Nevertheless, choosing initial conditions with the shocks close enough to each other determines well-defined initial conditions, which shows us the effect of the back reaction of the charge on the metric.  In a nutshell, the conclusions from these simulations are that (i) 
the effect of the charge on generic observables is relatively small, and (ii) this effect scales approximately linearly with $e^2$. We emphasize that these results are non-trivial since the charge density attains values well into what is expected to be the non-linear regime based on Eq.~\eq{lin}.

\begin{figure}[h!]
\begin{tabular}{cc}
\includegraphics[width=.452\textwidth]{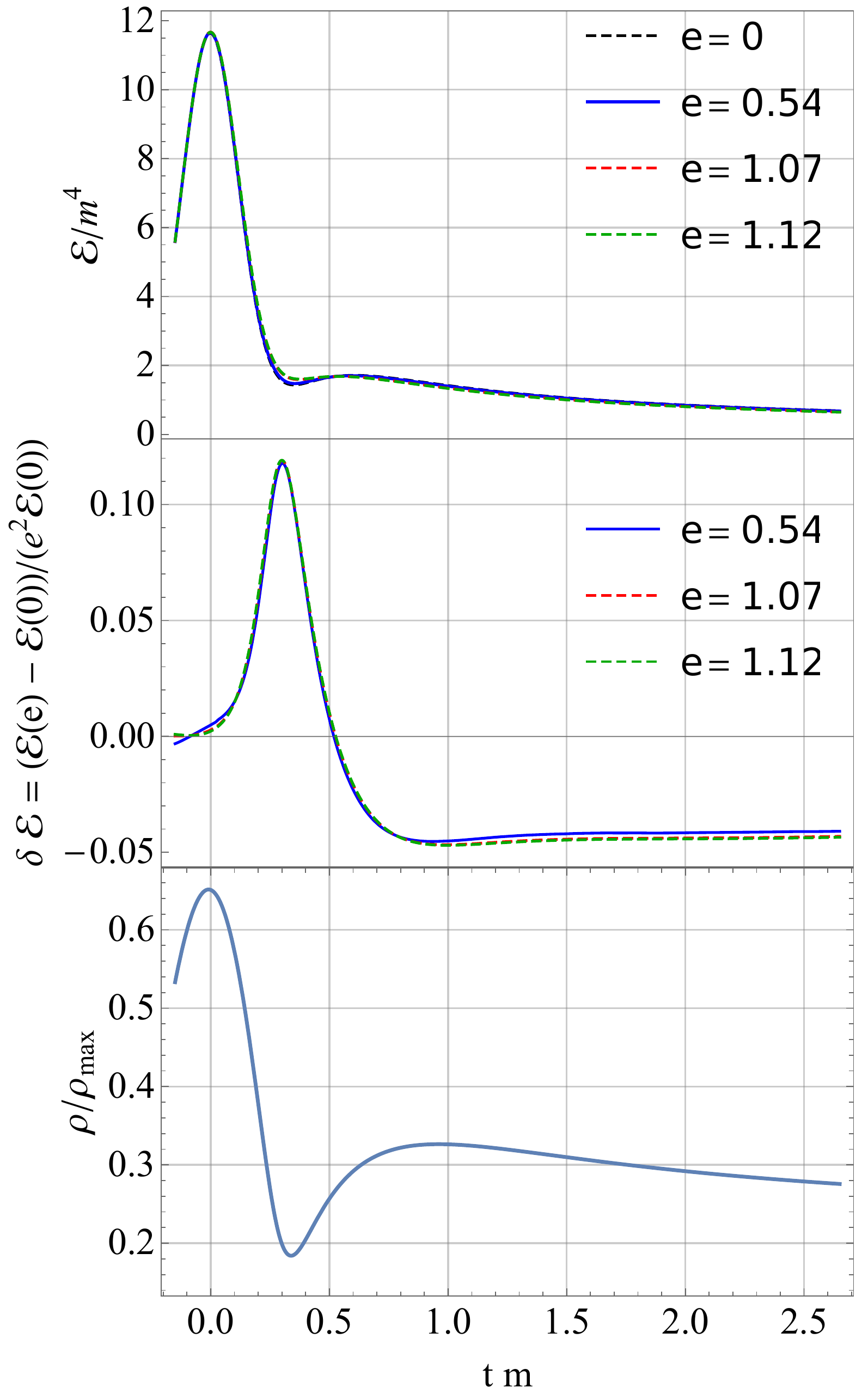} & \,\,
\includegraphics[width=.46\textwidth]{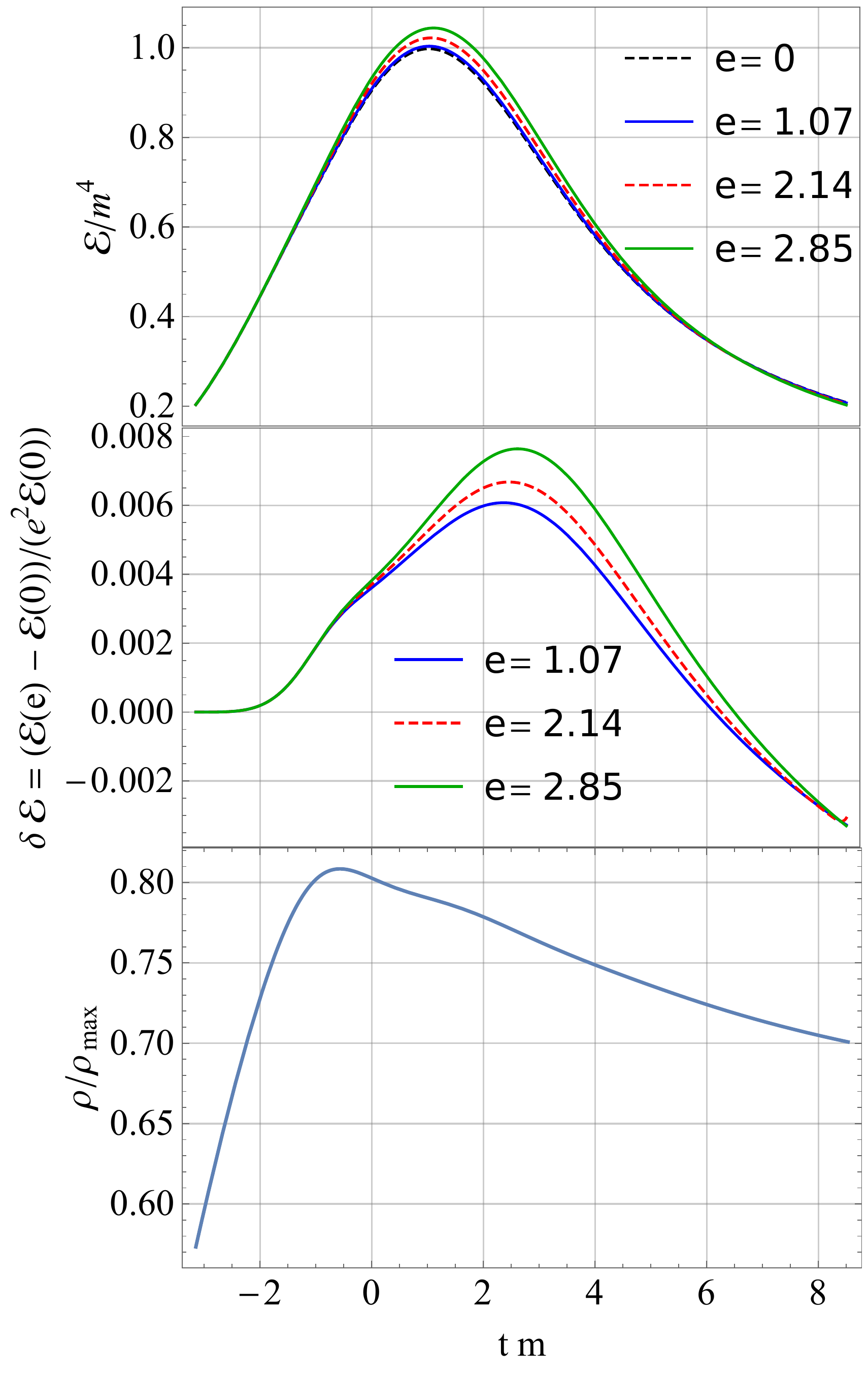}
\end{tabular}
\caption{\small Backreaction of the charge as a function of time. The first row shows the energy density at mid-rapidity as a function of time for collisions with different values of $e$ or, equivalently, different amounts of charge on the initial shocks, for thin  (left) and thick (right) shocks. The second row is the relative difference between a given collision and a collision with the same initial energy density but zero charge. The third row shows the charge density as function of time in units of $2 \eps^{3/4}/3$ at that time (see 
Eq.~\eq{ratio2}). 
}
\label{t}
\end{figure}
\begin{figure}[h!]
\begin{tabular}{cc}
\includegraphics[width=.452\textwidth]{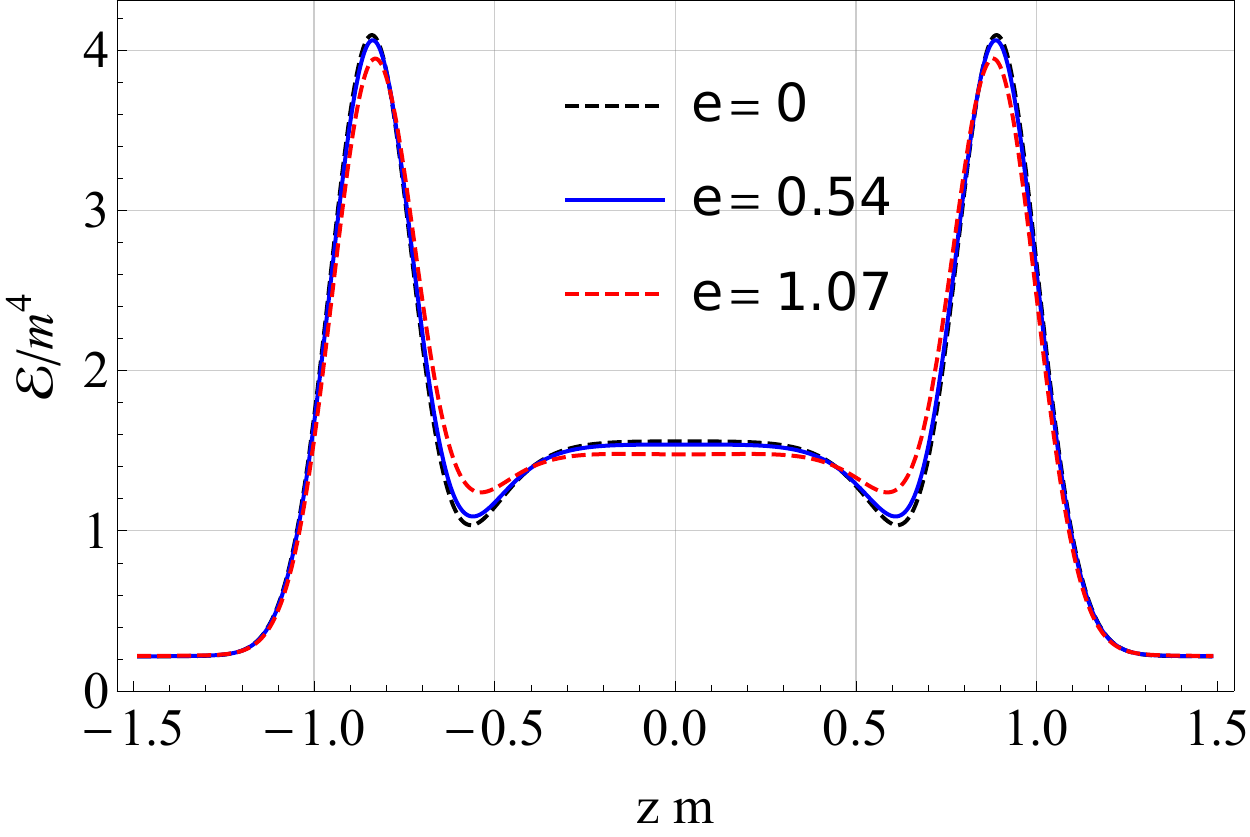} & \,\,
\includegraphics[width=.46\textwidth]{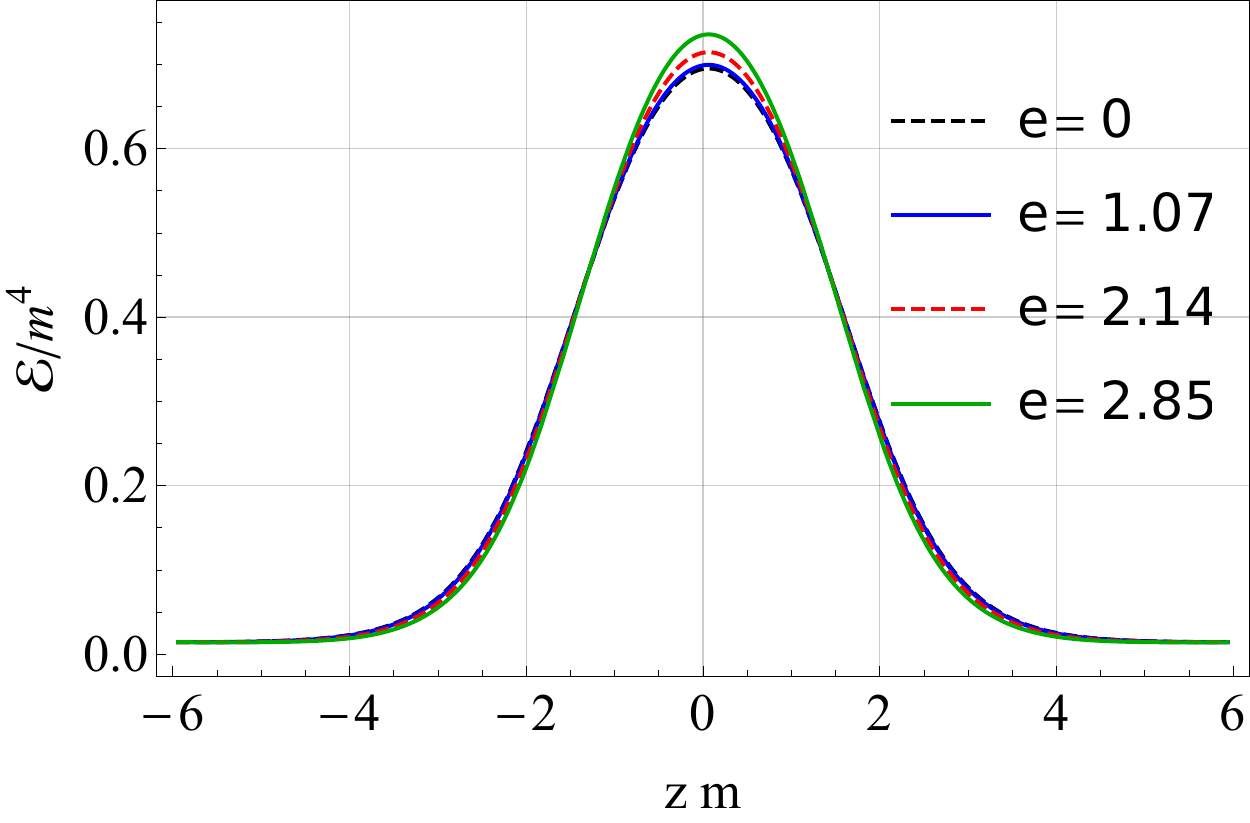}
\end{tabular}
\caption{\small Backreaction of the charge as a function of $z$. We show the energy density at $t=t_0$ as a function of $z$ for collisions with different values of $e$ or, equivalently, different amounts of charge on the initial shocks, for thin (left) and thick (right) shocks. The time $t_0$ is equal to $\thy$ for thin shocks and equal to the time at which the green curve in the second row of \fig{t} attains its maximum for thick shocks.}
\label{z}
\end{figure}

To illustrate these results, in \fig{t} we compare the energy density at mid-rapidity as a function of time for collisions with identical initial conditions but different values of $e$.  
The first row shows the energy density itself, whereas the second row shows the relative difference between a given collision and the $e=0$ collision,  normalized by $e^2$. The maximum value of the curves on the second row tells us that the maximum effect of the backreaction is about 
$0.12 \times 1.12^2 \simeq 15\%$ for thin shocks and 
$0.0075 \times 2.85^2 \simeq 6\%$ for thick shocks. The fact that the curves on the second row fall almost on top of each other for thin shocks (left) means that in this case the backreaction of the charge is almost exactly linear in $e^2$. For thick shocks the deviations from linearity are slightly larger. Note that in both cases the charge density as a function of time is not small. Indeed, for thin shocks it initially exceeds 60\% of the would-be maximum value according to the instantaneous energy density and drops to around 30\% at later times. For thick shocks the maximum exceeds 80\% and the curve drops to 70\% at later times. \fig{z} shows the same information as \fig{t} but as a function of $z$ at a constant time $t_0$. In the case of thin shocks we have chosen $t_0=\thy$, whereas in the case of thick shocks $t_0$ is the time at which the green curve in the second row of \fig{t} attains its maximum. 

As expected from this discussion, the effect of the charge density on the hydrodynamization time of the stress tensor is relatively small. For example, we find that for collisions with $m w = 0.1$ the hydrodynamization time in the charged case, with $e=1.07$, is 3\% shorter than in the neutral case. Instead, for collisions with $m w = 0.75$ the hydrodynamization time in the charged case, with $e=1.7$, is 6\% longer than in the neutral case.

\section{Discussion}

We have studied collisions of charged shocks  in AdS$_5$. Via the gauge/gravity duality, these are dual in the gauge theory to  collisions of lumps of energy carrying fixed amounts of baryon charge (per unit area). As discussed in \cite{Casalderrey-Solana:2013aba}, the dynamics of the stress tensor show qualitatively different features depending on the width of the colliding shocks. In this paper we have shown that similar qualitative differences also appear in the distribution of baryon charge after the collision. One of the main observations of \cite{Casalderrey-Solana:2013aba} was that, while thick shocks lead to a complete stopping of the incident energy followed by a subsequent hydrodynamic evolution, narrow shocks exhibit a transparent regime at early times in which the initial shocks cross each other depositing their energy gradually as time progresses.  In this paper we have shown that the baryon charge deposition exhibits identical behaviour, as illustrated in \fig{fig:snapshots}.

This transparent regime is transient, meaning that at a sufficiently late time the receding shock fragments are completely absorbed by the plasma. Nevertheless, the space-time rapidity distribution of baryon charge exhibits interesting diverse features for the two cases we have explored, as shown in \fig{fig:charge-rapidity}. At fixed proper times, both for thick and thin shocks the charge distribution is wider in rapidity than the energy distribution.  This means that, as rapidity grows, the plasma becomes more baryon rich, as illustrated by the increase in the chemical potential plotted in \fig{fig:chemical-potential}. However, for thin shocks the deposition of charge is wider than for thick shocks and it evolves much faster with time. Quite remarkably, for thin shocks the initial close-to-Gaussian charge distribution at early times, prior to hydrodynamisation, changes quickly to an almost-flat distribution in space-time rapidity which, within our limited numeric range,  hints at the formation of maxima at relatively large rapidity, $y>1$.  In the range of proper times and rapidities covered by our simulations, see \fig{fig:energy-rapidity}, these structures appear in regions in which the evolution is well described by hydrodynamics. However, the formation of these maxima may involve far-from-equilibrium dynamics that the hydrodynamic regions are in causal contact with. For thin shocks at very late times, the large rapidity region also develops local minima close to the edge of the rapidity coverage, which arise solely from the hydrodynamic evolution of the plasma.

\begin{figure}[t!]
\begin{tabular}{cc}
\includegraphics[width=.452\textwidth]{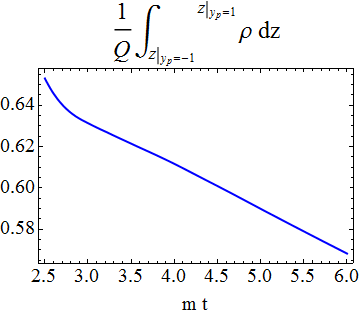} & \,\,
\includegraphics[width=.46\textwidth]{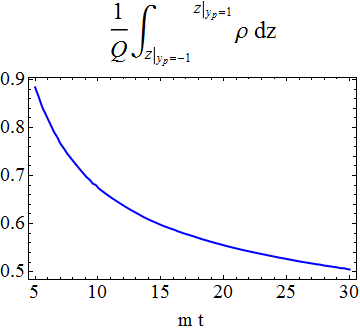}
\end{tabular}
\caption{\small We show the fraction of the charge (given in Eq.~(\ref{eq:charge})) in the plasma that has momentum rapidity $y_p = \tanh^{-1}(v_{\rm loc})$ smaller than 1.0 for thin (left) and thick (right) shocks. The plots start at the time that a fluid cell attains momentum rapidity 1.0. Clearly, a large fraction of the charge ends up at relatively small rapidities. Due to the hydrodynamic expansion this charge ends up at larger rapidity later on, which explains why the fraction decreases as a function of time.}
\label{chargeint}
\end{figure}

The space-time rapidity profile of the charge distribution hence indicates that collisions of shock waves lead to a significant stopping of the baryon charge of the incident projectiles. 
To best illustrate this point it is instructive to determine the fraction of the total charge (per unit area) of the incident shocks  that is deposited between momentum rapidity $y_p=-1$ and $y_p=1$, with $y_p={\rm arctanh} (v_z)$ and $v_z$ the velocity field in the collision direction.  This is determined by integrating the charge between the two points where the fluid velocity reaches $v_z = \tanh(1)\approx 0.76$, divided by the total charge. We show this quantity as a function of time in \fig{chargeint} for both thin and thick shocks. 

For thick shocks, the mid-rapidity charge fraction is bigger than for the thin shocks, which implies that the charge distribution is narrower in rapidity, as also illustrated in the right panel of \fig{fig:charge-rapidity}.
If the expansion of the system were exactly boost invariant, this quantity would remain constant in the hydrodynamic regime; therefore, the decrease of this fraction at later times is a consequence of non-boost invariant dynamics. Despite this decrease, 
this computation shows that more than 50\% of the charge carried by the initial shocks is deposited  within two units of rapidity at early times.\footnote{This quantity does not directly correspond to experimental measurements, as this requires a longer and more advanced hydrodynamic evolution and freeze-out. Nevertheless, for comparison, experimentally the fraction of baryon charge between rapidity -1.0 and 1.0 is approximately $37\%$ and $8.4\%$ for heavy ion collisions at $\sqrt{s_{\rm NN}}=17$ and $200$ GeV respectively \cite{Bearden:2003hx}.}

This strong baryon stopping is reminiscent of the behaviour of the net proton number in 
low-energy heavy ion collisions at $\sqrt{s}<20$ GeV. 
Indeed, 
heavy ion experiments performed at AGS \cite{Back:2000ru} and SPS \cite{Appelshauser:1998yb} found a large stopping of the baryon charge, with most of the net protons concentrated in the mid-rapidity region. As the energy of the beams increases, the rapidity width of the charge distribution increases, reducing the fraction of the total charge at mid-rapidity. Taking the width of our shocks as a proxy for the energy of the incident nuclei \cite{Casalderrey-Solana:2013aba}, this  trend  is in qualitative agreement with the behaviour of the charge distribution in our simulations.  This qualitative agreement indicates that shock wave collisions may provide a good framework to understand the hydrodynamisation of low- and moderate-energy heavy ion collisions, as those studied in the RHIC energy scan. 

In contrast, the dynamics of our simple holographic model does not seem to agree qualitatively with the distribution of net baryon number in heavy ion collisions at full RHIC or LHC energies. For those energies, the amount of net baryon number at mid-rapidity represents a small fraction of the total baryon number, and the distribution of baryon charge peaks a few units of rapidity away from the beam rapidity. Quite remarkably, as we have discussed, our thin shock simulations hint at the development of maxima at moderately large rapidity, which may be interpreted as  the onset of this non-monotonous behaviour. However, unlike in heavy ion collisions, most of the baryon charge is initially concentrated within a few units of rapidity.  

The discrepancy above is consistent with the difficulties in reproducing the LHC multiplicity rapidity profiles at high energies, as noted in \cite{vanderSchee:2015rta}. It may be due to the extreme simplicity of our model, which just evolves in the simplest holographic setting with the simplest Maxwell field possible, and for instance does not contain any matter in AdS. We are also working in a model that only describes the very first moments of a high energy collision, without treating a long hydrodynamic phase or freeze-out, and our model restricts to homogeneity in the transverse plane.
Naturally,  the
discrepancy may also point to a deeper difference between our holographic setup and the dynamics of QCD. 
Indeed, most of the baryonic charge of hadrons are carried by valence quarks, which also carry a large fraction of the full hadron momentum. In contrast, the shock waves posses structure functions  concentrated at small Bjorken $x$ \cite{Avsar:2009xf}. 
 Since processes able to reduce the rapidity of valence quarks  by a significant amount involve large momentum transfers, large rapidity shifts at high energies are suppressed as a consequence of asymptotic freedom. 
 The qualitative disagreement in the rapidity distribution of the charge is perhaps not surprising given the absence of this perturbative physics on the gravity side. Nevertheless, since the matter produced at mid-rapidities in a ultra-relativistic collision is soft, the interactions and generation of this matter may still be dominated by strong coupling processes. It would be interesting to develop hybrid approaches able to address these two separated regimes within the duality, perhaps along the lines of \cite{Casalderrey-Solana:2014bpa,Iancu:2014ava}.


\begin{acknowledgments}
We thank Michal Heller, Krishna Rajagopal and Paul Romatschke for
interesting discussions. JCS is a Royal Society University Research Fellow and was also supported by a Ram\'on y Cajal fellowship, by the Marie Curie Career Integration Grant FP7-PEOPLE-2012-GIG-333786 and by the Spanish MINECO through grant FPA2013-40360-ERC. We are also supported by grants MEC FPA2013-46570-C2-1-P, MEC FPA2013-46570-C2-2-P, MDM-2014-0369 of ICCUB, 2014-SGR-104, 2014-SGR-1474, CPAN CSD2007-00042 Consolider-Ingenio 2010, and ERC Starting Grant HoloLHC-306605. WS is supported by the U.S. Department of
Energy under grant Contract Number DE-SC0011090.

\end{acknowledgments}



\end{document}